\documentclass[twocolumn,aps,prl,groupedaddress]{revtex4}

\usepackage{graphicx}
\usepackage{color}
\usepackage{amsmath}
\usepackage{marvosym}
\usepackage{soul}
\usepackage{xcolor,cancel}
\usepackage{amssymb}
\usepackage{epstopdf}
\usepackage[normalem]{ulem}
\usepackage[shortlabels]{enumitem}
\setlist[enumerate, 1]{1\textsuperscript{o}}
\usepackage{setspace}

\newcommand{\be}{\begin{equation}}
\newcommand{\ee}{\end{equation}}
\newcommand{\bea}{\begin{eqnarray}}
\newcommand{\eea}{\end{eqnarray}}

\newcommand{\la}{\langle}
\newcommand{\ra}{\rangle}
\newcommand{\lb}{\left[}
\newcommand{\rb}{\right]}
\newcommand{\lp}{\left(}
\newcommand{\rp}{\right)}

\newcommand{\sgn}{{\rm sgn}\,}
\renewcommand{\Im}{{\rm Im}\,}
\renewcommand{\Re}{{\rm Re}\,}
\renewcommand{\vec}[1]{{\bf #1}}
\renewcommand{\epsilon}{\varepsilon}
\renewcommand{\tilde}{\widetilde}

\def\nn{\nonumber\\}

\def\dmag{|\vec d|}
\def\erad{\boldsymbol{\mathcal{E}}}

\begin{document}

\title{Intra-cell dynamics and cyclotron motion without magnetic field} 
\author{Eddwi H. Hasdeo$^1$, Alex J. Frenzel$^2$ and Justin C. W. Song$^{1,3}$}
\email{justinsong@ntu.edu.sg}
\affiliation{$^1$Institute of High Performance Computing, Agency for Science,
Technology, and Research, Singapore 138632\\ 
$^2$Department of Physics, University of California, San Diego, La
Jolla, California, USA  92039\\
$^3$Division of Physics and Applied Physics, Nanyang Technological University, Singapore 637371}
\begin{abstract}
Intra-cell motion endows rich non-trivial phenomena to a wide variety of quantum materials. The most prominent example is a transverse current in the absence of a magnetic field (i.e. the anomalous Hall effect). Here we show that, in addition to a dc Hall effect, anomalous Hall materials possess circulating currents and cyclotron motion without magnetic field. These are generated from the intricate wavefunction dynamics within the unit cell, and correspond to interband transitions (coherences) in much the same way that cyclotron resonances arise from inter-Landau level transitions in magneto-optics. Curiously, anomalous cyclotron motion exhibits an intrinsic decay in time (even in pristine materials) displaying a characteristic power law decay. This reveals an intrinsic dephasing similar to that of inhomogeneous broadening of spinors. Circulating currents can manifest as the emission of circularly polarized light pulses in response to incident linearly polarized (pulsed) electric field, and provide a direct means of interrogating the intra-unit-cell dynamics of quantum materials. 
\end{abstract} 

\maketitle

In crystals with a multi-basis unit cell (multiple atoms/orbitals), the weight of the wavefunction on each
basis plays a non-trivial role in the long wavelength dynamics of electrons. 
An illustrative example is graphene where a pseudo-spin arises from the relative wavefunction amplitude and phase on A and B sublattice sites within the unit cell: these define an intra-(unit) cell coordinate. 
This pseudo-spin is instrumental in a variety of Berry-phase related electronic phenomena that range from Klein
tunneling~\cite{katsnelson06} to a pinned zeroth Landau level~\cite{novoselov04,novoselov05,pkim05}.  
Non-trivial internal structure within the unit cell arises across a variety of multi-band quantum materials where geometric phases transform particle motion. 

Perhaps the most striking consequence of non-trivial intra-cell
coordinate motion is the anomalous Hall effect (AHE), in which
transverse currents can be induced by a longitudinal electric field
even in the absence of a magnetic
field~\cite{adams59,luttinger54,niu99,xiao10}. Although the average
position of electron wavepackets in a crystal is ambiguous and depends
on a gauge choice, intra-cell displacements, on the other hand, are
gauge invariant quantities. In anomalous Hall materials, the
non-trivial {\it structure} of intra-cell coordinates render the
components of the physical position operator/coordinate
non-commuting~\cite{adams59}.  This results in an anomalous velocity
(AV) transverse to the applied electric field in the semiclassical
equations of motion for particles~\cite{adams59,luttinger54,niu99,xiao10}.

\begin{figure}[t]
  \center
\includegraphics[width=8cm]{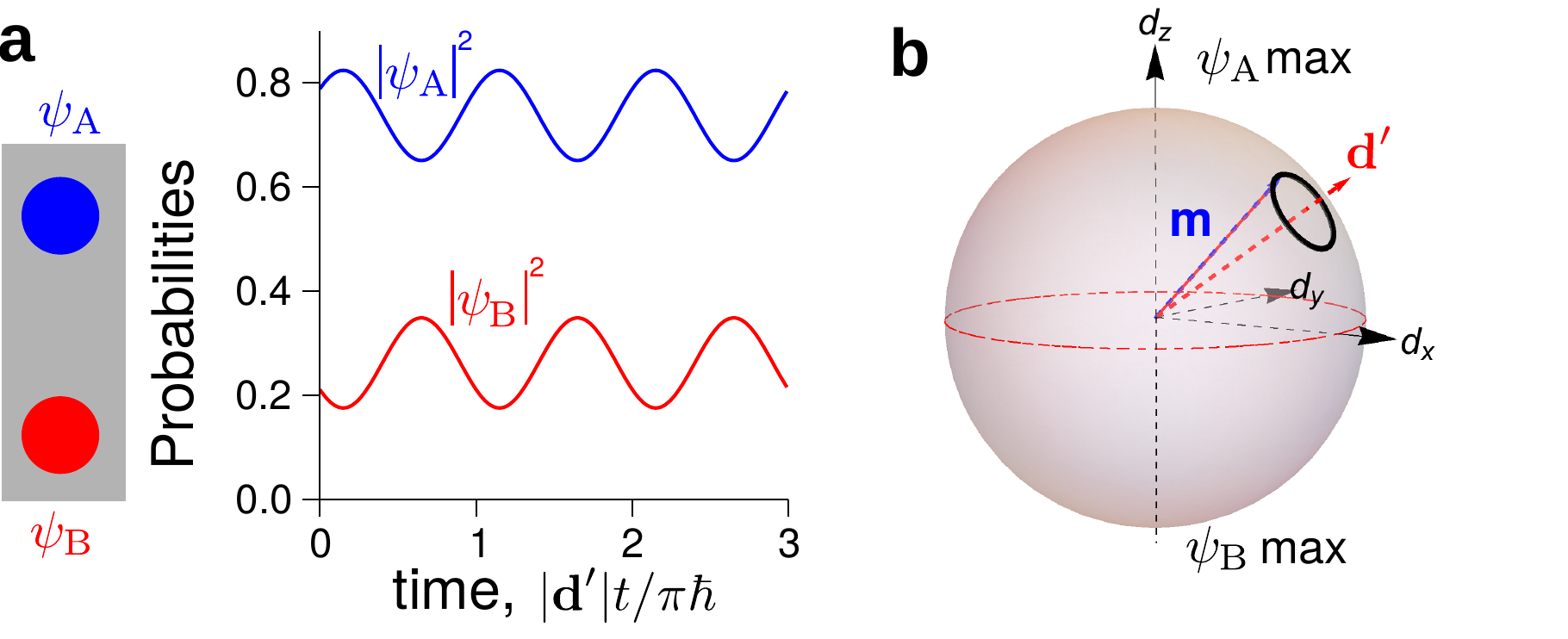}
\caption{\label{Fig1} {\bf Intra-cell dynamics.} {\bf a.} Weight of wave function $\psi_A,\psi_B$ on two sites of a unit cell can be modulated by an external electric field $E_x^\delta\delta(t)\hat{\vec x}$. Here $\vec d' = (d_x+evE_x^\delta,d_y,d_z)$. {\bf b.} The wave function is related to a pseudo-spinor, $\vec
  m= \la \psi(\vec
p)|\boldsymbol{\tau}|\psi(\vec p)\ra$, where $\boldsymbol{\tau}_{x,y,z}$ are Pauli matrices, see text. 
  Pseudo-spinor $\vec
  m$ (blue dashed line) is initially aligned to $\vec d$.  When an 
  electric field pulse $E_x^\delta\delta(t)\hat{\vec x}$ is applied, $\vec d$ is
  displaced to $\vec d'$. As a result, $\vec m$
  precesses around $\vec d'$. North (south) pole of Bloch sphere
  indicates a state where $|\psi_{\rm A}|^2=1$ ($|\psi_{\rm
    B}|^2=1$). Here we have used $d_x=d_y=d_z=\Delta$ and a small electric field pulse amplitude.}
\end{figure}

\begin{figure} [t]
  \center
\includegraphics[width=\columnwidth]{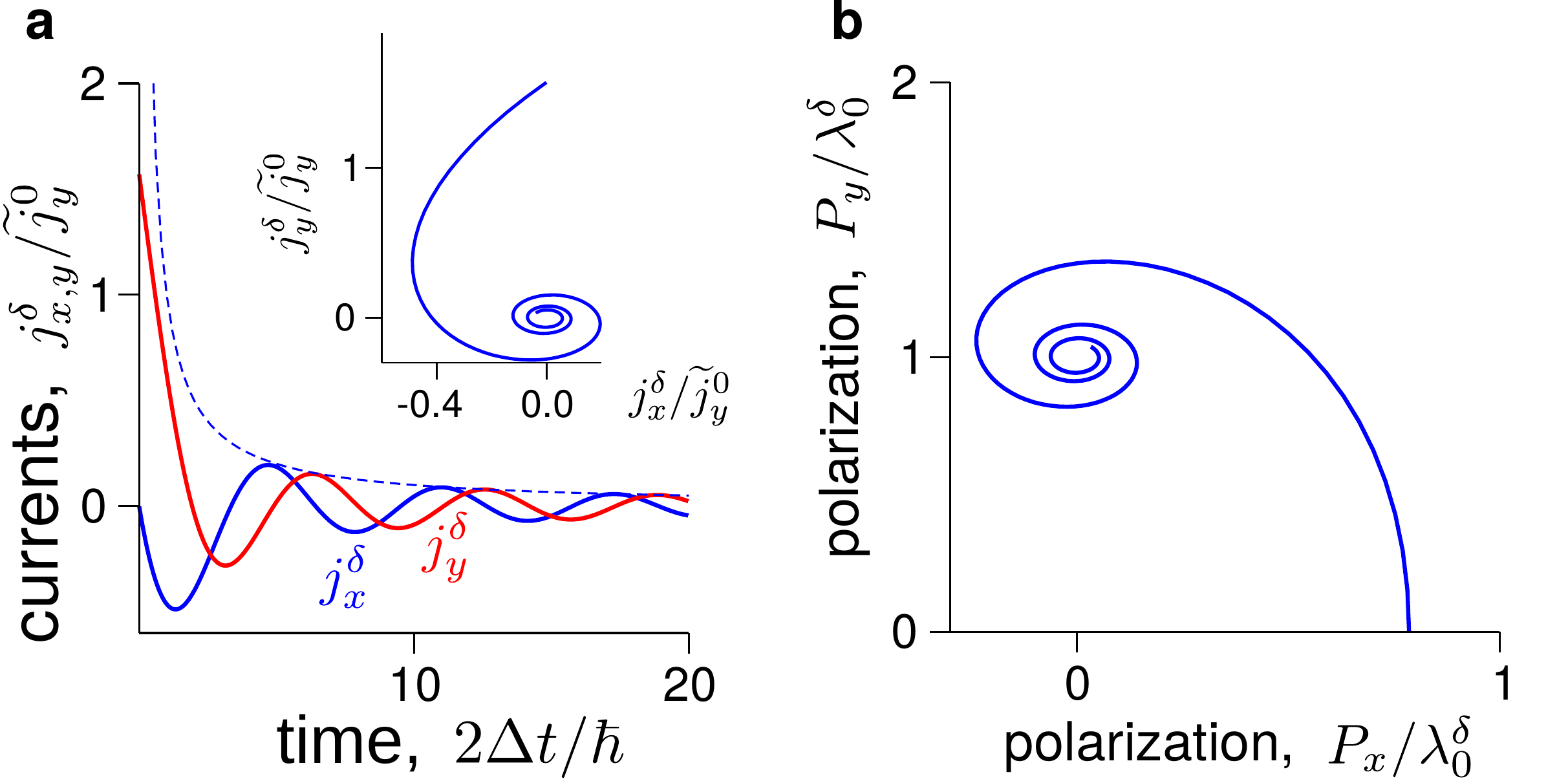}
\caption{\label{Fig2} {\bf Anomalous cyclotron motion}. {\bf a.} In the presence of a linearly polarized pulse, $E_x^\delta\delta(t)\hat{\vec x}$, 
  longitudinal current $ j_x^\delta$ (blue line) and Hall current $j_y^\delta$
  (red line) oscillate as a function of time. For
  $j_y^\delta$ plotted here, we have used $\chi = 1$. Dashed line indicates $1/t$
  decay. Here $\tilde j_y^0=E_x^\delta \sigma_H 2\Delta /\hbar$ and $\sigma_H=e^2/2h$. [Inset] $j_x^\delta$ vs $j_y^\delta$ are $\pi/2$ out-of-phase leading to a macroscopic circulating current. {\bf b.} Polarization exhibits an initial shift in $x$ due to the pulse. It subsequently spirals inward displaying a macroscopic (real-space) cyclotron motion of charge carriers in anomalous Hall materials. This is obtained from
  integrating $\vec j^\delta$ over time (see text) where
  $\lambda_0^\delta=\sigma_H E_x^\delta $.   }
\end{figure}

\begin{figure} [t]
  \center
\includegraphics[width=\columnwidth]{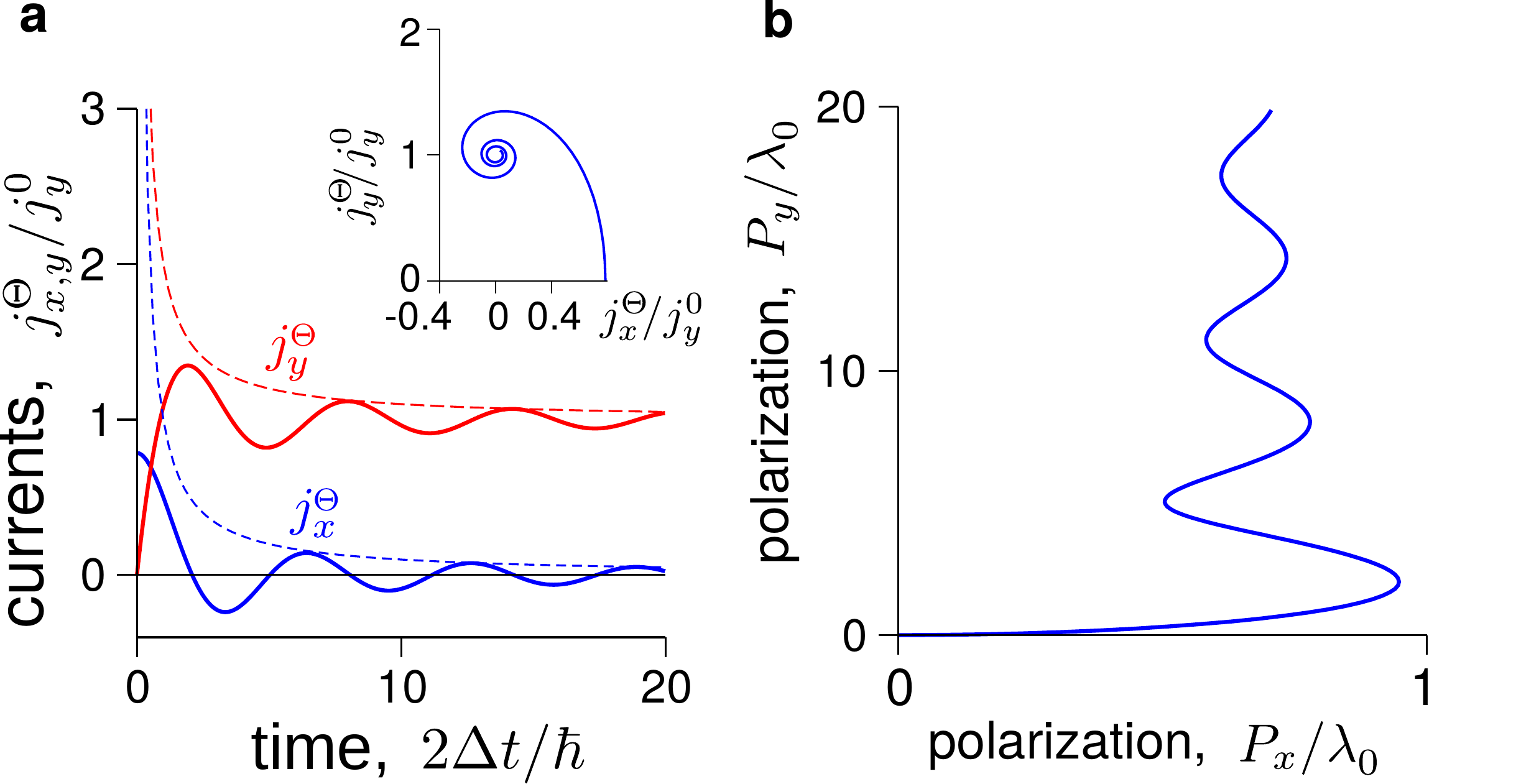}
\caption{\label{Fig3} {\bf Coexistence of anomalous velocity and anomalous cyclotron motion.} {\bf a.} In the presence of a step function and homogeneous (in space) 
  electric field $E_x^\Theta\hat{\vec x}$ at $t>0$,  
  longitudinal current $ j_x^\Theta$
  (blue line) and Hall current $j_y^\Theta$ (red line) oscillate in time with a $\pi/2$ phase shift. For
  $j_y^\Theta$, we have selected $\chi = 1$. Dashed lines indicate $1/t$
  decay. Here $j_y^0=\sigma_H E_x^\Theta $ with $\sigma_H=e^2/2h$. [Inset]
  Current density spirals about $(j_x^\Theta, j_y^\Theta ) = (0,1)j_y^0$ displaying ACM (circulating current) coexisting with AV ($j_y^\Theta = 1$). 
   {\bf b.} Polarization of carriers  obtained from integrating $\vec j^\Theta$ over time (see
  text). Here $\lambda_0=(\sigma_H/v)\lambda_c E_x^\Theta$ where
  $\lambda_c=\hbar v/2\Delta$ is the Compton wavelength.  }
\end{figure}

\begin{figure} [t]
  \center \includegraphics[width=8cm]{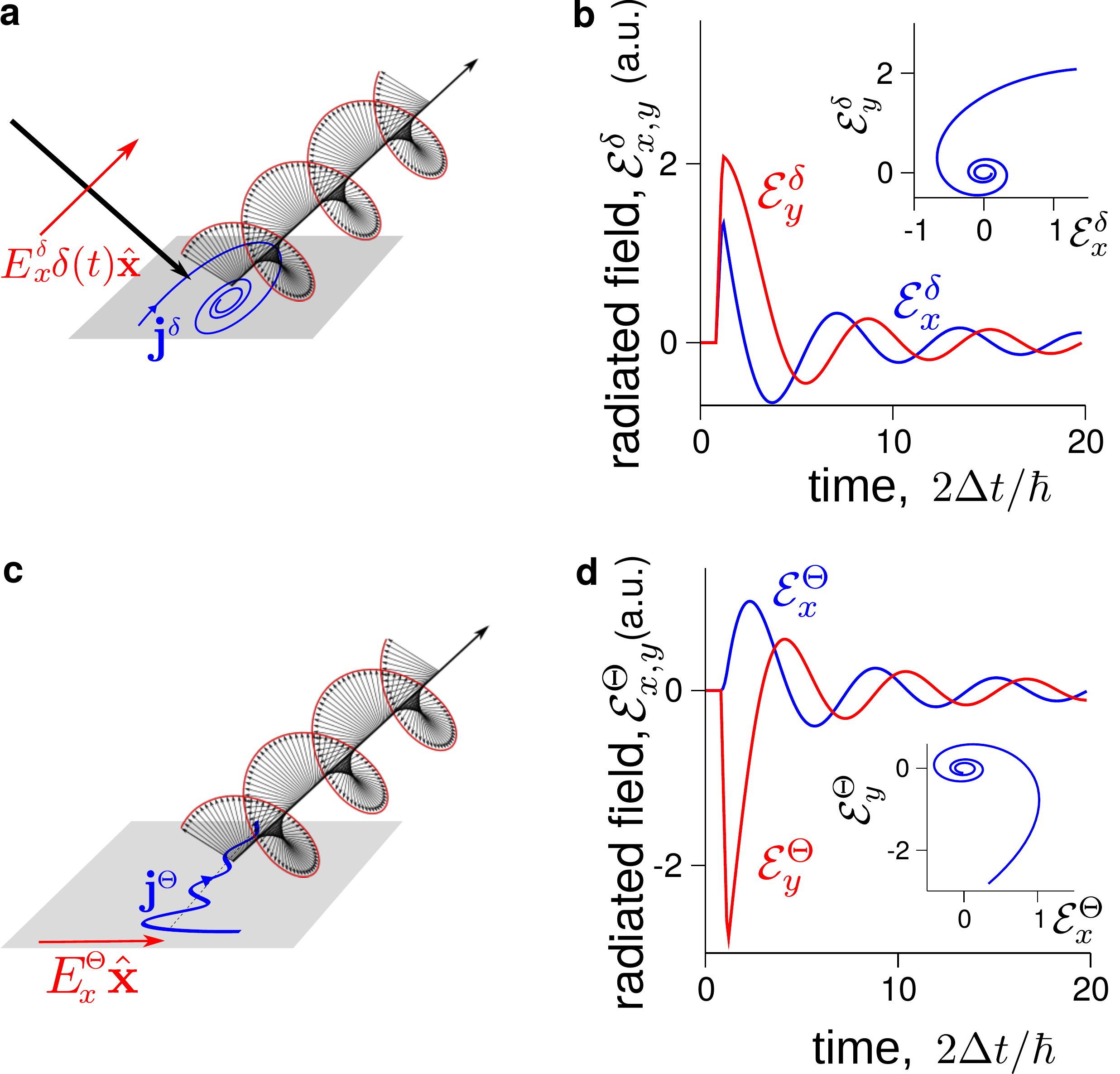}
\caption{\label{Fig4} {\bf Circularly polarized light emission without
    magnetic field.} {\bf a.} A linearly polarized pulse
  $E_x^\delta\delta(t)\hat{\vec x}$ yields a circulating current $\vec
  j$ and circularly polarized light emission.  {\bf b.} Radiated
  electric field from cyclotron current in $x$ (blue line) and $y$
  (red line) detected from a position $\vec r= (0,0,1)\ \mu {\rm
    m}$. We use $\chi=1$.  [Inset] Evolution of radiated
  electric field polarization. {\bf c.} Probing the intra-cell dynamics
  via emission of circular polarized light arising from a step function electric
  field $E_x^\Theta\Theta(t)\hat{\vec x}$. {\bf d.} Radiated electric field in
  $x$ (blue line) and $y$ (red line) from $\vec j^\Theta$ with the
  same parameters as {\bf b}. [Inset] Evolution of radiated electric
  field polarization. Jumps of $\erad$ at $t=|\vec r|/c$ are due to
  retardation. }
\end{figure}

Here, we show that in addition to a slow transverse drift arising from
AV, anomalous Hall materials exhibit unusual intra-cell dynamics with a
characteristic anomalous {\it cyclotron}-like motion without magnetic
field. For example, focussing on a prototypical anomalous Hall
material -- a broken time reversal symmetry (TRS) gapped Dirac system
with two sites per unit cell -- we find that the probability to find
an electron on either A or B sites (wavefunction square amplitude) can
naturally oscillate between the sites (Fig.~\ref{Fig1}a).  This
unusual intra-cell oscillator leads to a real-space cyclotron motion
(of the entire electron liquid) even when the valence band is fully
occupied. The anomalous cyclotron motion (ACM) can be activated by
either applying a short electric pulse (Fig.~\ref{Fig2}) or rapidly
turning on an electric field (Fig.~\ref{Fig3}).  Both approaches induce cyclotron
currents that can be probed via emission of circularly polarized light when
linearly polarized fields are incident on a sample~( Figs.~\ref{Fig4}).  

Similar to how {\it slow} guiding center motion (Hall drift) and {\it
  fast} cyclotron motion in the presence of a magnetic field are dual
to each other, the {\it slow} transverse drift from semiclassical AV
in anomalous Hall materials possesses an analogous duality to the {\it
  fast} ACM we describe here. This is because both ACM and AV arise
from the same origin --- a non-trivial quantum geometry in the Bloch
bands encoded in the intra-cell coordinate structure. To see this, we
recall that AV arises from the adiabatic motion of carriers projected
onto a single band. In contrast, ACM arises from the non-adiabatic dynamics
of carriers and corresponds to inter-band transitions between the
bands. This is similar to the relationship between magnetic field
induced guiding center motion and cyclotron motion which can be
understood as arising from the motion of carriers projected to a
single Landau level and virtual transitions between Landau levels
respectively. Much as cyclotron motion encodes the dynamical response found in magneto-optics, ACM is the physical manifestation of the dynamical response of anomalous Hall materials~\cite{mcd10,levitov11,brey14,rostami14,carbotte17}.

However, in contrast to conventional cyclotron motion in a magnetic
field, the ACM we discuss here exhibits intrinsic $1/t$ power-law
decay, even in a pristine and intrinsic system. This decay arises from
dephasing due to inhomogeneous broadening of the pseudo-spinors (in-momentum space). When electric field is
applied, pseudo-spinors in different states exhibit multi-frequency
Larmor precession (each pseudo-spinor precesses with a different
frequency and direction) thus creating dephasing. Nevertheless,
the response is peaked for transitions close to the band edges, and
displays a characteristic frequency that is tunable by the band gap in
a massive Dirac system.

We begin by considering a system with two sites in a unit cell, thus
having two bands. These two sites can correspond to two types of atoms
or orbitals within the unit cell; this two-band model is the simplest system exhibiting
a non-trivial intra-cell structure. We write the Hamiltonian as:
$\mathcal{H}=\boldsymbol{ \tau}\cdot \vec d(\vec p)$, where
$\boldsymbol{ \tau}_{x,y,z}$ are Pauli matrices that act on the
sublattice A and B sites, $\vec d(\vec p)$ describes the band
structure of the specific material, and $\vec p$ is the 
quasi-momentum. Intra-cell dynamics are encoded in the
pseudo-spinor $\vec m_{\vec p} =\la \psi(\vec
p)|\boldsymbol{\tau}|\psi(\vec p)\ra$, with $\psi(\vec p)=(\psi_A(\vec
p),\psi_B(\vec p))$ which compactly describes amplitudes and phases of the electron 
wavefunction on the A and B sites. The dynamics of $\vec m_{\vec p}$
thus represent intra-cell motion that can be tracked in the Bloch sphere
as illustrated in Fig.~\ref{Fig1}b. When $\vec m_{\vec p}$ is aligned
to the north (south) pole of the Bloch sphere, the wave function has
weight solely on A (B) sublattice.

Importantly, the pseudo-spinors obey the Bloch equation of motion
\be
\frac{d}{dt} \vec m_{\vec p} (t) =  \la \psi (\vec p) | \frac{1}{i\hbar}
[\boldsymbol{\tau}, \mathcal{H}] | \psi(\vec p) \ra= \frac{2}{\hbar}
\vec d(\vec p)\times \vec m_{\vec p}.
\label{eq:eom}
\ee 
At equilibrium (no applied electric field), 
$\vec m_{\vec p}= \vec m^{(0)}_{\vec p} =\pm \vec d (\vec p)/|\vec d (\vec p)|$ corresponding to the pseudo-spinors of conduction and valence band states respectively. At equilibrium, conduction (valence) band pseudo-spinors $\vec m^{(0)}_{\vec p}$ are parallel (anti-parallel) 
to $\vec d(\vec p)$ yielding $d\vec m_{\vec p}/dt$ that vanishes. When an external
electric field is applied, $\vec d$ is shifted by $\delta \vec d=\vec A \partial_\vec A \vec d(\vec p-e\vec
A/c)$ to linear order. Here $\vec A$ is the vector potential, related to electric
field by $\vec E=-\partial_t\vec A/c$. As a result, $\vec m_\vec p$ and $\vec
d'=\vec d(\vec p-e\vec A/c)$ are no longer aligned, turning on the dynamics of $\vec m_\vec p$. 

The rate of change of the pseudo-spinor, $d\vec m_{\vec p}/dt$, is perpendicular to $\vec m_{\vec p}$ (see
Eq.~\eqref{eq:eom}) which makes the pseudo-spinor $\vec m_{\vec p}$ Larmor precesses around
$\vec d'$ similar to spin dynamics (see Fig.~\ref{Fig1}
  b); this occurs naturally whenever $\vec m_{\vec p}$ is canted away from $\vec d (\vec p)$. This dynamics can be visualized as an oscillation of probability at each site (see Fig.~\ref{Fig1}a). As we will see, it is this intra-cell oscillation that leads to the unusual ACM we describe below.
In the following we will capture the dynamics by writing $\vec m_{\vec p} (t) = \vec m^{(0)}_{\vec p}+\delta\vec m_{\vec p}
(t)$. We note that Eq.~\eqref{eq:eom} can be nonlinear since $\delta \vec m$
and $\delta\vec d$ are both proportional to $\vec E$. In this work, however, we
focus on linear response while leaving non-linear effects as a subject
of further study.

To understand the intra-cell dynamics in anomalous Hall materials, we
specialize to TRS-broken gapped Dirac systems with $\vec d(\vec p)=(\chi
vp_x,vp_y,\Delta)$ where $\chi=\pm 1$ is the chirality, $v$ is the Dirac
velocity of electrons, and $\Delta$ is the energy difference between A
and B sublattice sites. Since TRS is broken, we will focus on just a single cone.  Although we focus on the simplest model,
equivalent qualitative results can be drawn for different systems such
as anomalous Hall ferromagnets,
topological insulators, and Weyl and Dirac semimetals [see also below for a discussion of the half-Bernevig-Hughes-Zhang (half-BHZ) model].

As we now argue, the Larmor-like intra-cell motion (Fig.~\ref{Fig1}b) persists in the macroscopic dynamics of the entire electron liquid and manifests as ACM. 
To see this, we excite the system with linearly polarized
light, $\vec E=E_0 e^{-i \omega t} \hat{\vec x}$. 
Collecting the linear
terms of Eq.~\eqref{eq:eom}, we obtain
\be
\frac{d}{dt} \delta \vec m_{\vec p} (t) =
\frac{2}{\hbar} \lp \vec d (\vec p)\times \delta \vec m_{\vec p} (t) + \delta \vec
d\times \vec m ^{(0)}_{\vec p}\rp,\label{eq:linear}
\ee
where $\delta \vec d=
i\chi e v E_0/\omega \hat{\vec x}$.  Here we have focussed on linear response, dropping the nonlinear term $\delta \vec
m_{\vec p}\times \delta \vec d (\vec p)$. We note that the dynamics of $\delta \vec m_\vec p$ is controlled by the internal
structure [the first term in the right hand side (RHS) of
  Eq.~\eqref{eq:linear}], as well as the external driving [the second term in RHS of Eq.~\eqref{eq:linear}]. The former gives rise to a precessional motion with frequency $2|\vec
d|/\hbar$ that corresponds to the oscillation frequency of virtual interband transitions. We note, parenthetically, that this can be related to the trembling motion or Zitterbewegung of Dirac fermions~\cite{katsnelson05}.
 
For a fully gapped system, with Fermi energy in the gap, the electrons are in the valence band. Solving for Fourier components $\delta\vec m \sim e^{-i\omega t}$ in Eq.~\eqref{eq:linear}, we obtain:
\bea
\delta m_x(\omega)&=&\frac{i \chi ev
  E_0(d_y^2+d_z^2)}{\omega \dmag ((\hbar\omega/2)^2-\dmag^2)},\nn 
\delta m_y (\omega) &=&\frac{-\chi ev E_0 \lp (\hbar\omega/2)
  d_z+id_xd_y\rp}{\omega \dmag ((\hbar\omega/2)^2-\dmag^2)},\nn 
\delta m_z(\omega) &=& \frac{\chi ev E_0\lp(\hbar\omega/2) d_y-i d_x
  d_z\rp}{\omega \dmag ((\hbar\omega/2)^2-\dmag^2)},
\label{eq:mxyzw}
\eea
where we have used $\vec m^{(0)}_{\vec p} =- \vec d (\vec p)/|\vec d (\vec p)|$ corresponding to the valence band. 

The current response of the entire electron liquid can be determined as $\vec j =e \sum_{\vec p} \la
\partial_{\vec p} \mathcal{H}\ra= ev (\chi \sum_{\vec p} \delta
m_x,\sum_\vec p \delta m_y,0)$, see supplementary information ({\bf SI}) for a full account.  Here the sum is taken over the entire valence band. 
We note that terms in
Eq.~\eqref{eq:mxyzw} that are odd in $d_x$ and $d_y$ cancel during
$\vec p$ integration. In the static limit $\omega\to 0$, the
non-vanishing integrand of $\delta m_y$ reproduces the
Berry curvature of gapped Dirac systems, $\chi v^2\hbar^2 d_z/(2\dmag^3)$; taking the
integral of $\delta m_y$ over $\vec p$, one can obtain the (half) quantized
Hall conductivity in the static limit expected from a single gapped Dirac cone per spin. 

To demonstrate the full response of intra-cell dynamics,
access to a broad range of frequency is needed.  This can be achieved, for example,
by an ultra short light pulse.  For simplicity, we first consider the form
$\vec E=E_x^\delta \delta(t)\hat{\vec x}$ (see {\bf SI} for a Gaussian profile for
a comparison). Using Eq.~\eqref{eq:mxyzw}, the current that develops at $t>0$ in response to the delta
function pulse is~\cite{supplement}: 
\begin{align}
  j_x^\delta(t) &=
  \frac{e^2}{4 h}\frac{2\Delta E_x^\delta}{\hbar} \bigg[\tau\lp {\rm Si}(\tau)-\frac{\pi}{2}\rp -\frac{{\rm sin}(\tau)}{\tau}\nn
    &+\cos(\tau)+\pi \delta(\tau)\bigg ],\nn
j_y^\delta(t) &= \frac{\chi e^2}{2h}\frac{2\Delta E_x^\delta}{\hbar} \lb\frac{\pi}{2}-{\rm
  Si}(\tau)\rb,\label{eq:deltacurrents} \end{align} 
  where $\tau=2\Delta
t/\hbar$, and ${\rm Si}(x)=\int_0^x dx' \sin(x')/x'$. 
Here $j_y^\delta$ captures the dynamical anomalous Hall motion. The oscillatory motion of both $j_x^\delta$ (blue) and $j_y^\delta$ (red) with $\chi=1$ are shown explicitly in 
Fig.~\ref{Fig2}a. This oscillation (finite frequency response) indicates the contribution from virtual interband transitions.

Crucially, the oscillatory response of $j_x^\delta$ and $j_y^\delta$
are displaced in phase by $\pi/2$ and characterizes the ACM in gapped
Dirac systems (Fig.~\ref{Fig2}a inset).  The $\pi/2$ phase lag of
$j_y^\delta$ vs $j_x^\delta$ is inherited from the intra-cell motion
that leads to the Larmor preccession of the individual spinors in
Fig.~\ref{Fig1}b.  Indeed, the plot $j_x^\delta$ vs $j_y^\delta$ in
the Inset of Fig.~\ref{Fig2}a shows a {\it circulating} current (ACM). The circulating current is centered at (0,0) because the delta-pulse electric field is (instantaneously) applied only at $t=0$.

To emphasize the real-space nature of ACM, we plot the change of polarization, $\Delta \vec P=\int_0^t
\vec j(t') dt'$ shown in Fig.~\ref{Fig2}b. This demonstrates how the average displacement for the carriers circulates in a cyclotron fashion reminiscent of that found in the presence of a Lorentz force. 
This features a characteristic spiral pattern, picking out a handedness (determined by $\chi$) that describes the sense of rotation; broken TRS is integral to ACM . 

We note, parenthetically, that before performing
cyclotron motion (ACM), the carriers exhibit a (global) shift in position. 
Because of $\delta(\tau)$ in $j_x^\delta$ [see Eq.~\eqref{eq:deltacurrents}], the electron liquid is initially displaced by
$\pi\lambda_0^\delta/4$ and returns to equilibrium in the long time limit.  Meanwhile in the $y$ direction, the electron liquid is polarized by an (electric-field-dependent) value $\chi
\lambda_0^\delta$ where $\lambda_0^\delta=\sigma_HE_x^\delta$ and $\sigma_H=e^2/2h$ is the dc Hall conductivity for a single gapped Dirac cone
per spin~\cite{supplement}. This can be understood by noting that, in the long time limit, polarization $\Delta \vec P$ from the delta function pulse is determined solely by the dc conductivity arising from intra-band contributions; we note that the inter-band contribution oscillates and its contribution to the polarization vanishes in the long time limit~\cite{supplement}. As a result, only a transverse polarization (arising from non-vanishing dc Hall conductivity) manifests at long times; longitudinal polarization diminishes to zero since dc longitudinal conductivity vanishes for a fully gapped system. 
This change of transverse polarization corresponds to the 
canting of $\vec d\to \vec d'$ in Fig.~\ref{Fig1}b and appears due to the 
instantaneous drift induced by the delta function pulse field.

In contrast to Larmor precession for a single $\vec p$
state in Fig.~\ref{Fig1}b, or to cyclotron motion in a magnetic field,
the macroscopic ACM found in $\vec j^\delta$ deteriorates, and follows a power-law $1/t$ decay
(see the dashed line in Fig.~\ref{Fig2}a).
This decay can be understood as a dephasing arising due to inhomogeneous broadening. Whenever the electric field is applied, each pseudo-spinor at state $\vec p$ precesses with different frequency and direction. At very short times, all the spinors oscillate at the same time. However, after sometime they begin to oscillate out of phase and dephase. 
As a result, ACM (the sum total dynamics of all the spinors) decays.
We note that this {\it intrinsic}
decay can be slower than extrinsic scattering processes which may cut
this slow-relaxation; nevertheless, the slow relaxation -- that occurs even in
the absence of extrinsic scattering -- is a Hallmark of the
non-trivial structure of the intra-cell coordinate dynamics.

Although the delta function pulse excites the entire frequency range ($\omega \in [0 , \infty]$), the temporal profile of 
$j_x^\delta$ and $j_y^\delta$ predominantly follows an oscillation 
frequency
$2 \Delta/\hbar$. This corresponds to the bandgap (i.e. direct interband transition between
band-edges) and is a physical manifestation of the sharp 
logarithmic resonance in $\Re[\sigma_{xy}(\omega)]$ and $\Im[\sigma_{xx}(\omega)]$ when $\omega$ matches
with energy gap $2\Delta$~\cite{supplement,mcd10}. This frequency further underscores the inter-band transition nature of ACM; the cyclotron-like intra-cell motion in these pulsed systems originate from {\it inter}-band wavefunction coherences induced by the pulse.

ACM (from inter-band coherence) can co-exist side-by-side with AV (from intra-band coherence). To see this, 
we apply a homogeneous electric field in space with a step function in time, $E_x^\Theta\Theta(t)\hat{\vec x}$, so as to get a dc response (corresponding to AV) as well as to incorporate a broad frequency range (to enable ACM discussed above).  
Using Eq.~(\ref{eq:mxyzw}), we obtain the full current response at $t>0$~\cite{supplement}: 
\begin{align}
j^\Theta_x (t) &=
\frac{e^2 E_x^\Theta}{8h}
\bigg[ (\tau^2-2)\lp {\rm Si}(\tau)-\frac{\pi}{2}\rp \nn
  &+\tau\cos(\tau)+\sin(\tau)\bigg ], \nn
j_y^\Theta(t) &=
\frac{\chi e^2E_x^\Theta}{2h} \lb 1-\cos (\tau)+\tau \lp\frac{\pi}{2}- {\rm
  Si}(\tau)\rp\rb \label{eq:currents}
\end{align}
 The step function (in time) and homogeneous (in space) field induces an
oscillating response with frequency $2\Delta/\hbar$ in both $j_x^\Theta$ and $j_y^\Theta$
(blue and red lines, respectively in Fig.~\ref{Fig3}a). 

The anomalous Hall current $j_y^\Theta$ for $\chi=1$ (red line of
Fig.~\ref{Fig3}a) oscillates around $j_y^0=\sigma_HE_x^\Theta$. While $\sigma_H=e^2/2h$ arises from the intra-band part of the AHE, 
the full response we evaluate here modulates the Hall current and includes an inter-band contribution (that arises from the sharp turn-on).

Similar to ACM induced by a delta-function pulse (in
Eq.~(\ref{eq:deltacurrents}) above), $j_x^\Theta$ and $j_y^\Theta$
decay as $1/t$ (dashed lines of Fig.~\ref{Fig3}a) and exhibits a phase
lag of $\pi/2$. Plotting $j_x^\Theta$ vs $j_y^\Theta$ in the Inset of
Fig.~\ref{Fig3}a, we display how ACM spirals inwards. We note that at
long times this oscillation decays and approaches the value expected
for adiabatic dc transport (i.e. a drift from AV); this is shown by
the spiral that is centered around $j_y^\Theta =j_y^0$ indicating its
long-time behavior. This shifted spiral indicates how ACM (most pronounced at short
times) can co-exist with AV.

For the step-function $E_x^\Theta\Theta (t)$,
the predominant (real-space) displacement (polarization) is in the transverse direction ($y$ direction); it is accompanied by an oscillatory
motion in $x$ (Fig.~\ref{Fig3}b). 
The oscillatory motion comes from the fact that $j_x^\Theta$
oscillates around zero (Fig.~\ref{Fig3}a). We note, parenthetically, there also exists a 
parallel translation in $x$ due to a non-zero integral of $j_x^\Theta$ over 
time. This translation is proportional to the applied electric field and
$\Delta^{-1}$ as determined by a quantity
$\lambda_0=(\sigma_H/v)\lambda_c E_x^\Theta$ where $\lambda_c=\hbar
v/2\Delta$ is a Compton wavelength. For small gap $2\Delta=10$~meV, the Compton wavelength can be sizable,  
$\lambda_c=65$~nm. It corresponds to the build up of polarization along $x$. 

Intra-cell dynamics can be probed by radiation of electromagnetic (EM) waves from the anomalous cyclotron current. Interestingly, these EM waves are circularly polarized even though the input pulse is linearly polarized, as illustrated in Figs.~\ref{Fig4}a and c.
To illustrate this, we note that EM radiation from a current source is given by:
\be
\boldsymbol{\mathcal{E}}(\vec r,t)=\frac{-1}{c^2}\int \frac{d^3\vec r'}{|\vec
  r'-\vec r|}\frac{d\vec j \lp \vec r',t-\frac{|\vec r'-\vec
    r|}{c}\rp}{dt},\label{eq:radi}
\ee where $c$ is the speed of
light. For the light pulse $E_x^\delta\delta(t)\hat{\vec x}$, we use a convenient choice for the laser spot size
assuming it to be a circle with radius $ 1\ {\rm \mu m} $ for illustration. For the step function electric field $E_x^\Theta\Theta(t)\hat{\vec x}$, we assume homogeneous current density $\vec
j^\Theta(\vec r,t)=\vec j^\Theta(t)\delta(z)$ over an area of $1\ {\rm \mu m}^2$; other spot sizes can be chosen with no qualitative changes to the results we discuss below.  The radiated electromagnetic waves $\erad^\delta$ and $\erad^\Theta$ are evaluated
at a position $\vec r=1\ {\rm \mu m}\, \hat{\vec z}$ above the source as shown in
Figs.~\ref{Fig4}b and d, respectively. Both $\erad^\delta$ and $\erad^\Theta$ decay following a power law $1/t$ and are circularly polarized (see Inset of Figs.~\ref{Fig4}b and d). The jumps of $\erad$ at $t=|\vec r|/c$ are due to the effect of retardation.

\begin{figure} [t]
  \center
\includegraphics[width=8cm]{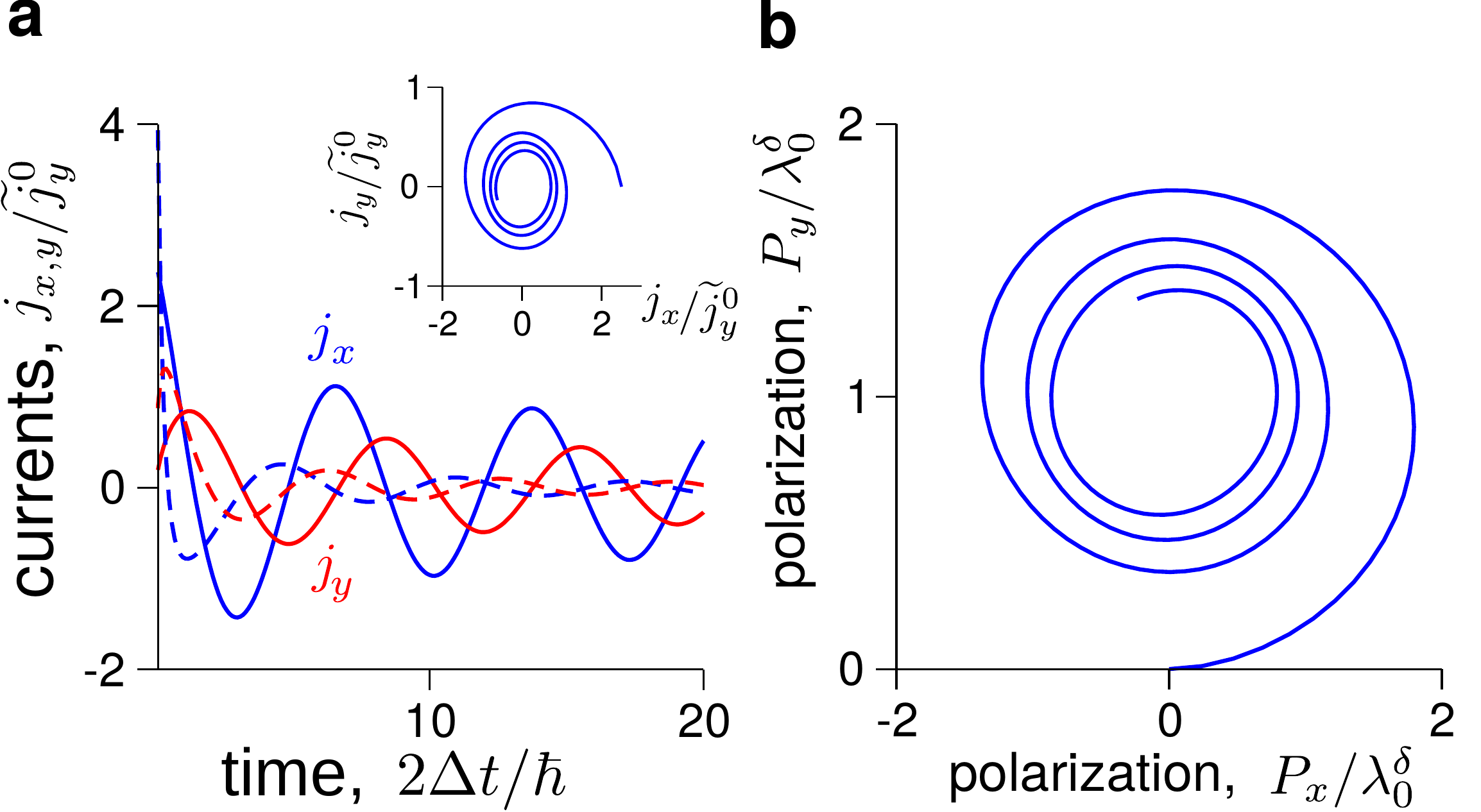}
\caption{\label{Fig5} {\bf ACM of half-BHZ systems}. {\bf a.} Current response of a half-BHZ system (see text) when excited by a delta function pulse $E_x^\delta\delta(t)\vec{\hat x}$. Longitudinal current $j_x$ (blue lines) and Hall current $j_y$ (red lines) are shown for two parameters
  $m^{-1}=v^2/\Delta$ (solid lines) and $m^{-1}=0.1v^2/\Delta$ (dashed lines) with $\chi=1$. [Inset] $j_x$ vs $j_y$ are $\pi/2$ out-of-phase leading to a macroscopic circulating current.  {\bf b.} Polarization of carriers show a a spiral-like character displaying a macroscopic (real-space) cyclotron motion. Here we used $m^{-1}=v^2/\Delta$ in Inset of {\bf a} and in panel {\bf b}.}
\end{figure}

ACM is a generic phenomena that arises in a variety of other anomalous Hall materials. As a further
example we consider
the half Bernevig-Hughes-Zhang (BHZ) Hamiltonian with
$\vec d^{\rm BHZ}=\lp \chi vp_x,vp_y,\Delta-m^{-1}(p_x^2+p_y^2)\rp$; this has TRS explicitly broken. Here $m^{-1}$ is the inverse mass parameter.
Using the same approach as above to tease out ACM, we apply a delta-function pulse
$E_x^\delta\delta(t)\vec{\hat x}$ and compare two cases of $m^{-1} =v^2/\Delta$ (solid lines) and $m^{-1}=0.1v^2/\Delta$ in Fig.~\ref{Fig5}. For both cases, $j_x$ and $j_y$ remain $\pi/2$ out-of-phase with each other. For small $m^{-1}$, the Hamiltonian tends to the same form as the Dirac system. As a result, the current response shows similar behaviour as compared to Fig.~\ref{Fig2}. For large $m^{-1}$, however, the amplitude of oscillation at long times is larger and lasts longer (with a smaller decay constant) than the gapped Dirac system analyzed above. Further. we note that $m^{-1}$ also modifies the oscillation frequency. Plotting  $j_x$ vs $j_y$, we show circulating current that is long-lived exhibiting a sizable radius (Inset Fig.~\ref{Fig5}).  As a
result, for large $m^{-1}$, we expect thagt ACM in the BHZ-type systems is more pronounced
compared to that found in massive/gapped Dirac systems (cf. Fig.~\ref{Fig5}b and
Fig.~\ref{Fig3}b).

In summary, we demonstrate how {\it fast} intra-cell dynamics can be tracked through a pulsed non-adiabatic
excitation, manifesting in ACM 
without magnetic field. ACM can be found in a variety of anomalous Hall materials and display a number of unusual characteristics including an intrinsic power-law decay, and real-space charge displacements that follow a spiral-like trajectory. This can be easily tracked via circularly polarized emission induced by a linearly polarized pulsed excitation. Just as slow anomalous velocity of carriers characterizes the intra-band quantum geometry (e.g., Berry curvature) of anomalous Hall materials, ACM captures its inter-band coherences. Both of these are rooted in a non-trivial intra-unit-cell structure and constitute different aspects of Bloch band geometry; ACM provides a dynamical window into the inner (often hidden) dynamics within the unit cell.

 {\bf Acknowledgements ---} We are grateful for useful conversations with Giovanni Viganle, Mark Rudner, Li-kun Shi, and Yang Bo. This work was supported by the Singapore
 National Research Foundation (NRF) under NRF fellowship award NRF-NRFF2016-05, and the Nanyang Technological University Singapore through a Start-up grant. 

\bibliographystyle{aip}


\setcounter{equation}{0}
\setcounter{figure}{0}
\renewcommand{\theequation}{S-\arabic{equation}}
\renewcommand{\thefigure}{S-\arabic{figure}}
\onecolumngrid
\newpage
\begin{center}
  {\large{\bf Supplementary Information for ``Intracell dynamics and cyclotron motion without magnetic field''}}\\
  \vspace{4mm}
  
Eddwi H. Hasdeo$^1$, Alex J. Frenzel$^2$ and Justin C. W. Song$^{1,3}$\\
{\it $^1$Institute of High Performance Computing, Agency for Science,
Technology, and Research, Singapore 138632}\\ 
{\it $^2$Department of Physics, University of California, San Diego, La
Jolla, California, USA  92039}\\
{\it $^3$Division of Physics and Applied Physics, Nanyang Technological University, Singapore 637371}
\end{center}

\vspace{3mm}
{\bf A. Intracell dynamics} 
\vspace{3mm}

In the following section, we will discuss intracell dynamics and its relation to cyclotron motion without magnetic field (anomalous cyclotron motion) in detail. 
To obtain the dynamics of the wavefunction within the unit cell it is useful to define the pseudospinors
\be
\vec m_\vec p = \la \psi (\vec p) |\boldsymbol{\tau} | \psi (\vec p) \ra, \quad  \psi (\vec p) = (\psi_A (\vec p) ,\psi_B (\vec p) ), 
\ee
where $\boldsymbol{\tau}_{x,y,z}$ are Pauli matrices that act on the sub-lattice $A$ and $B$ sites, with $\psi_A, \psi_B \in C$ are the wavefunction values on each of the sites respectively, and $\vec p$ is momentum. 

We will model the low-energy excitations (described by $\psi (\vec p)$) around each valley via a $2+1$ massive Dirac Hamiltonian 
\be
\mathcal{H} = \boldsymbol{\tau} \cdot \vec{d}(\vec p), \quad \vec{d}
(\vec p) = ( \chi vp_x, vp_y , \Delta  ) 
\label{eq:hamiltonian}
\ee
where $\chi$ is the chirality, $v$ is the Dirac velocity of the particles, and $\Delta$ the energy difference between $A$ and $B$ sublattice sites. 
Importantly, the pseudo-spinors obey the (Bloch) equations of motion
\be
\frac{d\vec m_\vec p}{dt} =  \la \psi (\vec p) | \frac{1}{i\hbar} [\boldsymbol{\tau}, \mathcal{H}] | \psi(\vec p) \ra= -\frac{2}{\hbar}  \vec m_\vec p\times \vec d(\vec p).
\label{eq:eom}
\ee
We note that at equilibrium (no applied electric field), $\vec m_{\vec p}= \vec m^{(0)}_{\vec p} =\pm \vec d (\vec p)/|\vec d (\vec p)|$ corresponding to the conduction and valence band states respectively. These pseudo-spinors of the conduction (valence) band states [at equilibrium] are parallel (anti-parallel) to $\vec d(\vec p)$ yielding $d\vec m_{\vec p}/dt$ that vanishes. However, as we discuss below, once an electric field is applied $\vec d$ cants turning on the dynamics of $\vec m_{\vec p}$. 

Writing $\vec m_\vec p (t)=\vec m_\vec p^{(0)}+\delta \vec m_\vec p (t) $, we solve for the linear response of the system:
\be
\frac{d}{dt} \delta \vec m_\vec p = -\frac{2}{\hbar} \lp 
\delta \vec m_\vec p \times\vec d (\vec p) +  \vec m_\vec p ^{(0)}\times \delta \vec d\rp,\label{eq:linear}
\ee
where $\delta \vec d= \vec A \partial_\vec A \vec d(\vec p-e\vec A/c)$.
We have dropped the nonlinear terms $\delta \vec m_\vec p\times \delta \vec d$.
Without losing generality, we have also specialized to a linearly polarized electric field
$\vec E= E_x e^{-i\omega t} \hat{\vec x}$. Inverting Eq.~\eqref{eq:linear}, we obtain $\delta \vec m$ directly as
\be
\lp
\begin{array}{c}
 \delta m_x\\
 \delta m_y\\
 \delta m_z
\end {array}
\rp
=
\frac{i\chi ev E_x}{\omega f(\omega)}\lp 
\begin{array}{c c c}
d_x^2-(\hbar\omega/2)^2 & d_xd_y+id_z(\hbar\omega/2) & d_xd_z-id_y(\hbar\omega/2)\\
d_xd_y-id_z(\hbar\omega/2) & d_y^2-(\hbar\omega/2)^2 & d_yd_z+id_x(\hbar\omega/2)\\
d_xd_z+id_y(\hbar\omega/2) & d_yd_z-id_x(\hbar\omega/2) & d_z^2-(\hbar\omega/2)^2
\end{array}
\rp
\lp
\begin{array}{c}
 0\\
 d_z/\dmag\\
- d_y/\dmag
\end {array}
\rp,
\label{eq:suppeqmatrixdirac}
\ee 
where $f(\omega)=i(\hbar\omega/2)\lp(\hbar\omega/2)^2-\dmag^2\rp$. In obtaining the above, we have specialized to the dynamics valence band electrons, since the system we focus on in the main text is fully gapped with Fermi energy in the gap. In so doing we have used $\vec m^{(0)}_{\vec p} =- \vec d (\vec p)/|\vec d (\vec p)|$ corresponding to the spinors valence band. Writing out \eqref{eq:suppeqmatrixdirac} explicitly we have
\bea
\delta m_x&=&\frac{i\chi e v  E_x(d_z^2+d_y^2)}{ g(\omega)}\nn
\delta m_y&=&\frac{-\chi e v E_x\lp (\hbar\omega/2) d_z+id_xd_y\rp}{g(\omega)}\nn
\delta m_z&=&\frac{\chi e v E_x\lp (\hbar\omega/2) d_y-id_xd_z\rp}{g(\omega)},
\eea
where $g(\omega)=\omega\dmag\lp(\hbar\omega/2)^2-\dmag^2\rp$.

\vspace{4mm}
\newpage
{\bf B. Longitudinal current}

\vspace{3mm}

We now proceed to obtain the current dynamics, first focussing on frequency space. Longitudinal current of the entire electron liquid can be obtained directly from the pseudo-spinors above as 
\be
 j_x(\omega)=e\sum_\vec p \la\partial \mathcal{H}/\partial p_x \ra=\chi  e v \sum_\vec p \delta  m_x,
 \ee
where the expectation value is taken for the valence band states by specifying $\vec m^{(0)}_\vec p=-\vec d(\vec p)/|\vec d(\vec p)|$ (see above) since we focus on a gapped Dirac system with Fermi energy in the gap. By using $\vec j(\omega)=\boldsymbol \sigma(\omega) \vec E(\omega)$, 
we write the longitudinal conductivity as
\be
\sigma_{xx}= \chi ev\sum_\vec p \frac{-i\chi
  ev(d_z^2+d_y^2)}{\omega\dmag \lp (\hbar\omega/2)^2-\dmag^2 \rp}.\label{eq:sigxx}
\ee
The real and imaginary parts of longitudinal conductivity can be obtained from the usual substitution $\omega\to\omega+i\eta$ where $\eta\to 0^+$.
We note that the denominator of Eq.~\eqref{eq:sigxx} can be decomposed into
\be
\frac{1}{(\Omega+i\eta)^2-D^2}=\frac{1}{2D}\lp\frac{1}{\Omega+i\eta-D}-\frac{1}{\Omega+i\eta+D}\rp.
\ee
Using Sokhotski-Plemelj relation
\be
{\rm lim}_{\eta\to 0+}\frac{1}{x\pm i\eta}={\rm p.v}\ \frac{1}{x}\mp i\pi\delta(x),
\ee
we directly obtain the real part of $\sigma_{xx}$~[1--4] as
\bea
\Re \lb \sigma_{xx}(\omega)\rb&=&\frac{i4e^2v^2}{(2\pi\hbar)^2}\int_p\frac{p dp d\theta(d_z^2+d_y^2)}{\dmag\omega}\frac{-i\pi\lp\delta(\hbar\omega-2\dmag)-\delta(\hbar\omega+2\dmag)\rp}{4\dmag},\nn
&=&\frac{\pi e^2}{8h}
\frac{\Omega^2+1}{\Omega^2}\Theta(|\Omega|-1),\quad\Omega=\frac{\hbar\omega}{2\Delta}.
\label{eq:realpartconductivityxx}
\eea

We note that the explicit functional form of the real part of the conductivity alone is enough to determine the $t$-dynamics of current we focus on in the main text (see description below). However, for completeness, we also display the imaginary part of $\sigma_{xx}$ using this formulation. This can be obtained directly obtained from Eq.~\eqref{eq:realpartconductivityxx} via the Kramers-Kronig relations~[1--4]:
\bea
\Im\lb\sigma_{xx}(\omega)\rb&=& -\frac{1}{\pi}\int_{-\infty}^{\infty}
\frac{\Re \lb\sigma_{xx}(\omega')\rb}{\omega'-\omega} d\omega',\nn
&=& \frac{-e^2}{4h}\lb \lp 1+\frac{1}{\Omega^2}\rp
\mathcal{S} (\Omega^{-1})-\frac{1}{\Omega}\rb, \quad 
\mathcal{S}(x)=\bigg\{
\begin{array}{c}
{\rm acoth}(x),\quad x>1\\
{\rm atanh}(x),\quad x\le 1
\end{array}, \label{eq:imsigma}
\eea
where we have used the identity $\int dx \frac{x^2+1}{x^2(x-x_0)}= \mathrm {ln} |x-x_0| \lp 1 + x_0^{-2} \rp-x_0^{-2} \mathrm{ln} |x|+ (x x_0)^{-1}$, and the relation $\mathcal{S}(x)=\frac{1}{2}\mathrm{ln} \left|\frac{x+1}{x-1}\right|$.

We now proceed to the $t$-dynamics central to the anomalous cyclotron motion discussed in the main text. 
Of particular interest, is the current dynamics that ensues after the application of a delta function pulse $\vec E(t) = E_x^\delta\delta(t)\hat {\vec x}$. This can be written as
\bea
\vec j^\delta (t)&=&\frac{E_x^\delta\Theta(t)}{2\pi}\int_{-\infty}^{\infty}d\omega \left\{\Re\lb
\boldsymbol \sigma(\omega)\rb + i \Im \lb \boldsymbol \sigma(\omega)\rb \right \} e^{-i\omega t},\nn
&=& \frac{E_x^\delta\Theta(t)}{2\pi}\left\{\int_{-\infty}^{\infty}d\omega  e^{-i\omega t} \Re\lb
\boldsymbol \sigma (\omega)\rb + i \lp \frac{-1}{\pi}\rp\int_{-\infty}^{\infty} d\omega'\Re \lb \boldsymbol \sigma (\omega')\rb \int_{-\infty}^\infty d\omega\frac{e^{-i\omega t}}{\omega'-\omega}\right\} ,
\nn
&=& \frac{2E_x^\delta\Theta(t)}{2\pi}\int_{-\infty}^{\infty}d\omega e^{-i\omega t}\Re[\boldsymbol \sigma (\omega)].\label{eq:jdelta} 
\eea
In obtaining Eq.~\eqref{eq:jdelta}, we transformed $\Im[\sigma_{xx}]$
and expressed it in terms of $\Re[\sigma_{xx}]$ via the Kramers-Kronig
relation. We have also used the identity $\int_{-\infty}^\infty du
e^{-iut}/u =-i\pi \sgn (t)$. 

Employing Eq.~\eqref{eq:jdelta} and using Eq.~\eqref{eq:realpartconductivityxx} we have
\bea
j_x^\delta(t)&=&\frac{ e^2}{4 h}\frac{2\Delta E_x^\delta\Theta(t)}{\hbar} \lb \cos(\tau)+\tau\lp {\rm
  Si}(\tau)-\frac{\pi}{2}\rp-{\rm sinc} (\tau)+\pi\delta(\tau)\rb, \quad \tau=\frac{2\Delta t}{\hbar}, \label{eq:jxdelta}
\eea

The first two terms in Eq.~\eqref{eq:jxdelta} come from $\int_1^\infty d\Omega \cos (\Omega
\tau)/\Omega^2$, and we have used identities ${\rm Si}(x)=\int_0^x
dy\ \mathrm{sinc}( y)$ and $\int_0^\infty dy\ \mathrm{sinc}( y)=
\frac{\pi}{2}$, with $\mathrm{sinc}( y)= \sin (y)/y$. The two last
terms come from the Fourier transform of the inverse top-hat function $\Theta
(|\Omega|-1)$.  The inverse top-hat function ensures a zero $\Re[\sigma_{xx}]$ when $|\hbar \omega|$ is below the gap size $2\Delta$ and a finite value when $|\hbar\omega| >2\Delta$. Because $\Theta
(|\Omega|-1)$ is an even function, the integral yields $\rm sinc (\tau)$.  
The delta function $\delta(\tau)$ appears
due to finite Dirac conductivity ($\Re[\sigma_{xx}]$) at
$\omega\to \infty$. This delta
function $\delta(\tau)$ is responsible for the initial shift of the average carrier
position as we integrate $j_x(t)$ over time (see Fig. 2b in the main
text). We note that the functions $\cos(\tau) + \tau (\mathrm{Si}(\tau)-\pi/2)$ as well as $-{\rm sinc} (\tau)$ oscillate
 with  envelopes that decay as $1/\tau$. As a result, $j_x^\delta$ is an oscillating function that  decays as $1/\tau$.

Next, we discuss the case of a homogeneous electric field (in space) and step function in time $\vec E(t)= E_x^\Theta \Theta (t)\hat{\vec x}$. The current response $j_x^\Theta$ is given by
\bea
j_x^\Theta (t) &=& \frac{E_x^\Theta}{2\pi}\int_{0}^t dt' \int_{-\infty}^{\infty} d\omega \left\{ \Re\lb\sigma_{xx}(\omega)\rb +i\Im\lb\sigma_{xx}(\omega)\rb\right\} e^{-i\omega (t-t')}\nn
&=&  \frac{2E_x^\Theta}{2\pi} \Theta (t) \int_{-\infty}^{\infty}d\omega
\frac{i(e^{-i\omega t}-1)}{\omega}\Re[\sigma_{xx}(\omega)],\nn
&=&\frac{e^2E_x^\Theta\Theta (t)}{8
  h}\lb\sin(\tau)+\tau\lp \cos(\tau)+ \tau \lp {\rm
  Si}(\tau)-\frac{\pi}{2}\rp \rp - 2 \lp {\rm
  Si}(\tau)-\frac{\pi}{2}\rp\rb  ,\quad \tau=\frac{2\Delta t}{\hbar},
\label{eq:jdc}
\eea
where $\tau$ is a dimensionless time. Here we have expressed $\Im[\sigma_{xx}]$ into $\Re[\sigma_{xx}]$ via Kramers-Kronig relation similar to Eq.~\eqref{eq:jdelta}. The first two terms come from integration by parts,  $2\int_1^\infty d\Omega \frac{\sin(\Omega\tau)}{\Omega^3}=\sin (\tau) + \tau \int_1^\infty \frac{\cos(\Omega \tau)}{\Omega^2}$ while the last terms come from usual sine integral $\int_\tau^\infty dx\ \mathrm{sinc}(x)=\frac{\pi}{2} - \mathrm{Si} (\tau)$. The combined contribution of the first two terms as well as the last term are oscillating functions with envelopes that decay as $1/\tau$. As a result, $j_x^\Theta$ is an oscillating function that  decays as $1/\tau$.

\vspace{3mm}

{\bf C. Anomalous Hall current}

\vspace{3mm}

We now turn to the anomalous Hall current $j_y$ which can be evaluated as 
\be
j_y(\omega)=ev\sum_\vec p \delta m_y,
\ee
where the summation \vec p is taken for the valence band states. From this we obtain the real part of anomalous Hall conductivity~[1--3],
\bea
\Re[\sigma_{xy}(\omega)]&=&\frac{-\chi e^2v^2}{\omega}\sum_\vec p \frac{(\hbar\omega/2) d_z+id_xd_y}{((\hbar\omega/2)^2-\dmag^2)\dmag},\nn
&=& \frac{\chi e^2}{2h}\frac{1}{\Omega}\mathcal{S}(\Omega^{-1}),\quad 
\mathcal{S}(x)=\bigg\{
\begin{array}{c}
{\rm acoth}(x),\quad x>1\\
{\rm atanh}(x),\quad x\le 1\end{array},
\eea
where $\Omega=\frac{\hbar\omega}{2\Delta}$ and we have used the identity $\int_1^\infty \frac{dx}{a^2-x^2}=-\frac{1}{a}\mathcal{S} (a^{-1})$.  At $\omega\to 0$, $\sigma_{xy}$ shows a half-quantized value $e^2/2h$ for a single gapped Dirac cone. This transverse motion is non-dissipative since it arises from an undergap intraband contribution. At finite $\omega$, $\sigma_{xy}(\omega)$ is peaked at $\hbar\omega=2\Delta$ showing a logarithmic resonance; it subsequently decreases for $\hbar\omega> 2\Delta$. This resonant behaviour is a signature of virtual interband transition.

For the imaginary part $\Im[\sigma_{xy}]$, we substitute $\omega\to \omega+i\eta$ and follow the same procedure as  $\Re[\sigma_{xx}]$ to obtain~[1--3]:
\be
\Im[\sigma_{xy}(\omega)]=\frac{-\chi e^2 \pi}{4 h} \frac{1}{\Omega} \Theta (|\Omega|-1).\label{eq:imsigmaxy}
\ee
The imaginary part of the anomalous Hall conductivity is dissipative response from interband transitions. We note that the real and imaginary parts of the Hall conductivity are related via Kramers-Kronig relations.

Next, we calculate the  Hall response current for a delta function pulse $\vec E(t) = E_x^\delta\delta(t)\hat {\vec x}$. From the previous analysis in Eq.~\eqref{eq:jdelta} above, we note that the contributions to the charge current $t$-dynamics from the real and imaginary parts of Hall conductivity are equal.
As a result, we obtain 
\bea
j_y^\delta(t)&=&2\frac{E_x^\delta\Theta(t)}{2\pi}\int_{-\infty}^{\infty}d\omega 
\Re[\sigma_{xy}(\omega)] e^{-i\omega t}\nn
&=&\frac{\chi e^2}{2h}\frac{2\Delta E_x^\delta\Theta(t)}{\hbar}\lb
\frac{\pi}{2}-{\rm Si} (\tau)\rb,\label{eq:jydelta}
\eea
where we have used $\int_{-\infty}^{\infty} d\omega \frac{e^{-i\omega t}}{\omega^2-a^2}=-\frac{\pi}{a}\sin (a t)$ and the usual sine integral $\int_\tau^\infty dx\ \mathrm{sinc}(x)=\frac{\pi}{2} - \mathrm{Si} (\tau)$. $j_y^\delta(t)$ is an oscillating function that decays as $1/t$. The current predominantly oscillates with frequency determined from the gap size $\omega =2\Delta/\hbar$ due to the logarithmic resonance of ${\rm Re} [\sigma_{xy} (\omega)]$.

We proceed to obtain the Hall current for homogeneous electric field in space and step function in time $E_x^\Theta\Theta(t)\hat{\vec x}$,
\bea
j_y^\Theta(t) &=& \frac{2E_x^\Theta}{2\pi}\int_{0}^t dt' \int_{-\infty}^{\infty} d\omega \Re[\sigma_{xy}(\omega)]e^{-i\omega (t-t')},\nn
&=& \frac{\chi e^2 E_x^\Theta}{2 h}\Theta (t) \lp 1 -\tau \lp {\rm
  Si}(\tau)-\frac{\pi}{2}\rp -\cos (\tau) \rp,\quad \tau=\frac{2\Delta t}{\hbar}.
\eea
We have used an identity $\int_{-\infty}^{\infty}d\omega \frac{i(e^{-i\omega t}-1)}{\omega(\omega-a)(\omega+a)}=-\frac{\pi}{a^2} \lp\cos (at)-1\rp$ and $\int_1^\infty d\Omega \cos (\Omega \tau)/\Omega^2=\cos(\tau)+\tau\lp {\rm Si}(\tau)-\frac{\pi}{2}\rp$ as obtained previously in Eq.~\eqref{eq:jxdelta}. $j_y^\Theta$ shows an oscillating behavior that decays as $1/t$ from an interband contribution. The constant dc Hall current $e^2/2h E_x^\Theta$ arises from an intraband contribution. 

\vspace{3mm}

{\bf D. Change of polarization in a delta function pulse}

\vspace{3mm}

Electric polarization is defined by  $\partial_t \vec P =
 \vec j$. This can be seen by defining the polarization as
$\nabla\cdot \vec P=-\rho$, where $\rho$ is charge density and taking
the continuity equation $\nabla\cdot \vec j +\partial_t \rho =0$. In the case of a delta function pulse, the
whole time integral of current response (polarization) is
related to the dc conductivity. This can be seen directly from 
\bea
\int dt \vec j^\delta (t)
&=&\frac{2 E_x^\delta}{2\pi}\int_{-\infty}^{\infty} dt \Theta (t)
\int_{-\infty}^{\infty} d\omega \Re[ \boldsymbol\sigma(\omega)]
e^{-i\omega t},
\nn
&=& E_x^\delta \Re[\boldsymbol\sigma (0)],\label{eq:timej} \eea
In obtaining this, we have used the Fourier transform $\int_{-\infty}^{\infty}dt e^{-i\omega t}\Theta(t) =\pi \delta (\omega) - i/\omega$ and noted that only the delta
function $\delta(\omega)$ survives as $\Re[\boldsymbol \sigma(\omega)]/\omega$ is
antisymmetric function that vanishes upon integration of
$\omega$. Equation~\eqref{eq:timej} suggests that, for a fully gapped system, only the Hall shift
 (polarization in $y$) is finite due to a quantized dc Hall conductivity $\sigma_{xy}(\omega = 0)$.

\vspace{3mm}

{\bf E. Half Bernevig-Hughes-Zhang (BHZ) model}

\vspace{3mm}

The half BHZ Hamiltonian has a $\vec d$ vector that follows $\vec d=(\chi v p_x,v p_y,\Delta -m^{-1}
(p_x^2+p_y^2))$, where $m^{-1}$ is the inverse of the mass term.  With applied electric field in the $x$ direction, $p_x\to
p_x+ieE_x e^{-i\omega t}/\omega$, we have $\delta d_x= i\chi e v E_x /\omega$ and $\delta d_z=-2i\chi e  m^{-1} p_x E_x/\omega$. We have dropped the nonlinear terms in $\vec E$ since we focus on linear response. Solving the linear equation~\eqref{eq:linear}, we obtain
\be
\lp 
\begin{array}{c c c}
-i\hbar\omega/2 & d_z & -d_y\\
-d_z & -i\hbar\omega/2 & d_x\\
d_y & -d_x & -i\hbar\omega/2
\end{array}
\rp
\lp
\begin{array}{c}
 \delta m_x\\
 \delta m_y\\
 \delta m_z
\end {array}
\rp
=\lp
\begin{array}{c}
-\delta d_z m^{(0)}_y \\
 \delta d_z m^{(0)}_x-\delta d_xm^{(0)}_z\\
 \delta d_x m^{(0)}_y
\end {array}
\rp.
\ee 
Inverting the matrix yields
\bea
\delta m_x &=& \frac{i \chi e  E_x\lp v d_y^2+v d_z^2+2m^{-1} d_xp_x d_z-2im^{-1}
  d_yp_x (\hbar\omega/2)\rp }{g(\omega)}\nn
\delta m_y &=& \frac{-\chi e  E_x
  \lp v(\hbar\omega/2)d_z+i v d_xd_y+2m^{-1} d_xp_x(\hbar\omega/2)-2im^{-1}  d_y d_z p_x\rp}{g(\omega)}\nn
\delta m_z &=& \frac{-\chi e  E_x \lp-v(\hbar\omega/2)d_y+iv d_x d_z+2im^{-1}
  d_x^2 p_x+ 2i m^{-1} d_y^2 p_x\rp}{g(\omega)},
\eea 
where $g(\omega)=\omega\dmag ((\hbar\omega/2)^2-\dmag^2)$. 

We now isolate the terms that are even in $\vec p$ since that they do not vanish upon integration over $\vec p$: 
\bea
\delta m_x &=& \frac{i \chi e  E_x\lp v d_y^2+v d_z^2+2m^{-1} d_xp_x d_z \rp}{g(\omega)}\nn
\delta m_y &=& \frac{-\chi e  E_x(\hbar\omega/2)
  \lp v d_z+2m^{-1} d_xp_x\rp}{g(\omega)}\nn
\delta m_z &=& 0
\eea

From this we can evaluate the real part of $\sigma_{xx}$,
\bea
\Re\lb\sigma_{xx} (\omega)\rb
&=&\frac{e^2v^2}{ h\hbar\omega }\int_p\frac{p dp
  (\Delta^2+v^2p^2/2+m^{-1} p^2\Delta)}{\dmag}\frac{\pi\lp\delta(\hbar\omega-2\dmag)-\delta(\hbar\omega+2\dmag)\rp}{\dmag},
\eea
where the integral of $\vec p$ is taken for valence band states.
The longitudinal current for the delta function electric field can be expressed as
\bea
j_x^{\delta}(t) &=&  \frac{2E_x^\delta}{2\pi} \Theta (t) \int_{-\infty}^{\infty}e^{-i\omega t}\Re\lb\sigma_{xx}(\omega)\rb d\omega,\nn
&=&\frac{e^2}{2h}\frac{E_x^\delta 2\Delta \Theta(t)}{ \hbar}\int_0^\infty P dP \lp
1+P^2\lp\frac{1}{2}+\tilde\beta\rp\rp \frac{\cos (D\tau)}{D^3}\label{eq:jxdeltabhz},
\eea
where $\tilde \beta = m^{-1}\Delta/v^2$, $P=vp/\Delta$, and
$D=\dmag/\Delta$, and $D= (P^2+(1-\tilde\beta P^2)^2)$. We then obtain $j_x\delta$ by integrating Eq.~\eqref{eq:jxdelta} numerically.

Next, we move to the Hall current/transverse response. To analyze this we write the real part of Hall conductivity as
\bea
\Re[\sigma_{xy}(\omega)]&=&\frac{-\chi e^2v^2}{\omega}\sum_\vec p
\frac{(\hbar\omega/2)( d_z+2m^{-1} p_x^2) }{((\hbar\omega/2)^2-\dmag^2)\dmag},\nn
&=&\frac{-\chi e^2}{2h}\int_0^\infty PdP
\frac{1}{(\Omega^2-D^2)D},\quad \Omega=\frac{\hbar\omega}{2\Delta},
\eea
where the summation of $\vec p$ is taken for the valence band states. We note that $m^{-1}$ dependence on $\Re[\sigma_{xy}]$ is encoded inside $D$. Finally, we use this to obtain the Hall current numerically via the integral
\bea
j_y^\delta(t)&=&\frac{2E_x^\delta}{2\pi}\Theta(t)\int_{-\infty}^{\infty}e^{-i\omega t}
\Re[\sigma_{xy}(\omega)] d\omega,\nn
&=&\lp\frac{\chi e^2}{2h}\frac{E_x^\delta 2\Delta \Theta(t)}{\hbar}\rp\int_0^\infty P dP \lp\frac{\sin(D\tau)}{D^2}\rp,
\eea
where we have used $\int_{-\infty}^{\infty}d\omega \frac{e^{-i\omega t}}{\omega^2-a^2}=-\frac{\pi}{a} \sin(a t)$. The numerically evaluated current $t$-dynamics are discussed in the main text. 

\newpage
\vspace{5mm}

{\bf F. Excitation of Dirac fermion with a Gaussian pulse}

\vspace{3mm}
We have shown that the contribution of the real part of the conductivity tensor is sufficient to describe the $t-$dynamics of the current response in the case of delta function [Eq.~\eqref{eq:jdelta}] and step function fields [Eq.~\eqref{eq:jdc}]. In this section, we use the fact that 
this is also true for a general shape of electric field. This can be seen directly by following the analysis in Eq.~\eqref{eq:jdelta} above to write the complex $\boldsymbol \sigma$ only in terms of $\Re[\boldsymbol \sigma]$ through the Kramers-Kronig relation. We obtain:
\bea
\vec j(t) &=& \frac{1}{2\pi}\int_{-\infty}^t dt' \int_{-\infty}^{\infty}d\omega \boldsymbol \sigma(\omega) e^{-i\omega(t-t')}\vec E(t'),\nn
&=& \frac{1}{2\pi}\int_{-\infty}^t dt' \vec E(t')\int_{-\infty}^{\infty}d\omega 2\Re[\boldsymbol \sigma(\omega)] e^{-i\omega(t-t')}.\label{eq:general}
\eea
In obtaining the above result, we have noted that the $\omega$ integral in Eq.~\eqref{eq:general} is similar to Eq.~\eqref{eq:jdelta} by substitution of $t\to t-t'$.
\begin{figure} [t]
  \center
\includegraphics[width=8cm]{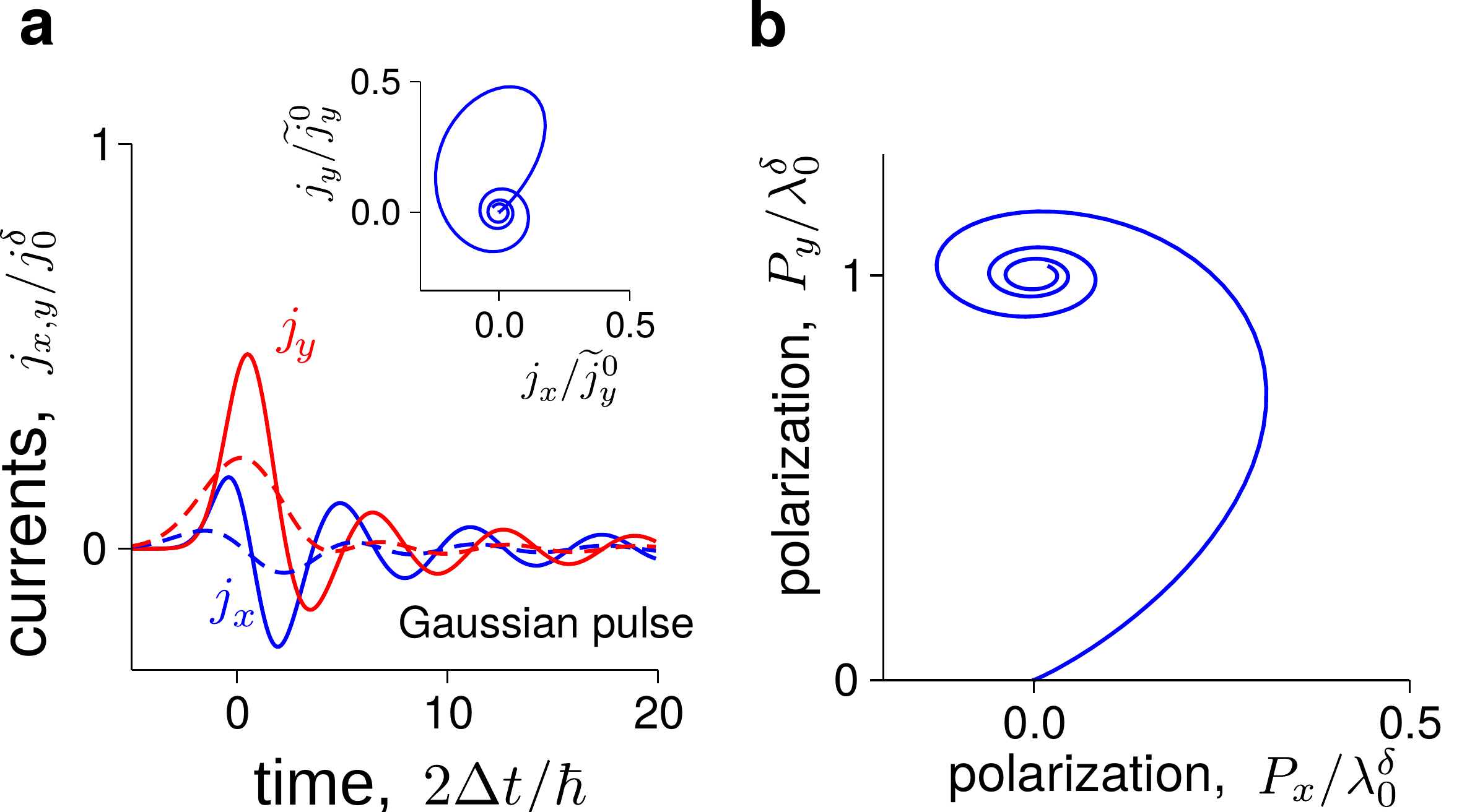}
\caption{\label{SFig1} {\bf ACM with Gaussian pulse} {\bf a} Current response for gapped Dirac
  systems excited by Gaussian pulse shows oscillation with $\pi/2$ phase difference between $j_x$ and $j_y$. For $j_y$ shown in this plot, we have used $\chi = 1$. We used the Gaussian pulse widths $\delta
  t=\hbar/2\Delta$ (solid lines) and $\delta t=\hbar/\Delta$ (dashed
  lines).  [Inset] Due to $\pi/2$ phase difference of $j_x$ and $j_y$, the current circulates exhibiting ACM. We used
  $\delta t=\hbar/2\Delta$. {\bf b} Polarization (evolution over time) from the same Gaussian pulse also exhibits anomalous cyclotron
  motion in real space. In plotting panel b we used Gaussian pulse width $\delta t=\hbar/2\Delta$. }
\end{figure}

As a simple sanity check, we test this for a single mode drive $\vec E(t)=\vec E_0 e^{-i\omega_0 t}$. The response should be $\vec j(t)=\boldsymbol\sigma(\omega_0)\vec E_0 e^{-i\omega_0 t}$. Applying Eq.~\eqref{eq:general} we obtain
\bea
\vec j(t)&=& \frac{1}{\pi}\int_{-\infty}^t dt' \vec E_0 e^{-i\omega_0t'+\eta t'}\int_{-\infty}^{\infty}d\omega \Re[\boldsymbol \sigma(\omega)] e^{-i\omega(t-t')}\nn
&=&\frac{\vec E_0}{\pi}\int_{-\infty}^{\infty}d\omega \Re[\boldsymbol \sigma(\omega)] \frac{-i e^{-i\omega_0t}}{(\omega_0-\omega)+i\eta}\nn
&=&\frac{\vec E_0e^{-i\omega_0t}}{\pi}\int_{-\infty}^{\infty}d\omega \Re[\boldsymbol \sigma(\omega)] \lp \frac{-i }{(\omega_0-\omega)}+\pi \delta(\omega-\omega_0)\rp\nn
&=&\vec E_0e^{-i\omega_0t}\lp i \Im[\boldsymbol \sigma(\omega_0)]+\Re[\boldsymbol \sigma(\omega_0)] \rp,
\eea
yielding the expected result. In obtaining the above, we have taken $\lim \eta\to 0$ and used the Kramers-Kronig relation to obtain $\Im [\boldsymbol \sigma]$ in the first term.

We now simulate ACM in a gapped Dirac system with a more realistic pulse: an electric field pulse with Gaussian profile $E_x=
\frac{E_x^\delta}{\sqrt{2\pi \delta t^2} }\exp[-t^2/2\delta t ^2]$ where $\delta t$ is the pulse width. We plug this profile
into Eq.~\eqref{eq:general} and obtain the current response numerically
in Fig.~\ref{SFig1}. In this plot, we show current response for two cases of $\delta t = \hbar/2\Delta$ (solid lines) and $\delta t = \hbar/\Delta$ (dashed lines). We find that as long as the pulse width $\delta t$ is smaller than or  about the same  as $\hbar/2\Delta$, the current response is
qualitatively similar to that of delta function pulse (compare Fig.~\ref{SFig1}a in the supplement with Fig. 2a of the main text) with similar frequency oscillation; it also exhibits a similar power law decay. Increasing the
pulse width obviously reduces the amplitude of oscillation (cf. solid [small pulse width]
vs dashed [large pulse width] lines). Therefore, the ACM $t-$dynamics will not be seen
in a single frequency continuous wave excitation. Additionally, we note that for small pulse width, the current response is insensitive
to the central frequency of the Gaussian pulse.
Plotting $j_x$ vs $j_y$ in the inset of Fig.~\ref{SFig1}a shows circulating currents (ACM) due to the $\pi/2$ phase difference between $j_x$ and $j_y$. ACM turns on slowly due to finite width of the Gaussian pulse. In the long time limit; current is centered at $(0,0)$ similar to that of the delta function pulse (cf. inset of Fig.2a in the main text) since the driving field turns off at long times. 

In Fig.~\ref{SFig1}b we show the polarization of the carriers (integrated over the entire electron liquid). As expected, the charge displacement in the long time limit along $y$ reproduces that of delta function pulse (see Fig. 2b in the main text). This originates from intraband contribution to the Hall current. Along the $x$ direction, the carriers are displaced slowly at early times due to finite width of the Gaussian. At long times, the total change of polarization in $x$ vanishes since dc longitudinal conductivity is zero [see Eq.~\eqref{eq:timej}] similar to that found for the delta function electric field pulse discussed in the main etxt.

\vspace{3mm}
\begin{spacing}{0.5}
{\small
  \begin{enumerate}[{[1]}]
   \item W.-K. Tse and A.~H. MacDonald, Phys. Rev. Lett. {\bf 105}, 057401 (2010).
  
 \item M. Lasia and L. Brey, Phys. Rev. B {\bf 90}, 075417 (2014).

\item  H. Rostami and R. Asgari, Phys. Rev. B {\bf 89}, 115413 (2014).

    \item  T. Stauber, J. Phys. Condens. Matt. {\bf 26}, 123201 (2014).
\end{enumerate}}
\end{spacing}
\end{document}